\documentclass{aa}
\usepackage{graphicx}
\usepackage{natbib}
\usepackage{txfonts}
\usepackage{supertabular}

\begin{document}
\def\ab{$\sim$}
\def\frac{$''$\hspace*{-.1cm}}
\def\hi{\ion{H}{i}}
\def\h2{\ion{H}{ii}}
\def\cii{\ion{C}{ii}}
\def\oiii{\ion{O}{iii}}
\def\hb{H$\beta$}
\def\ha{H$\alpha$}
\def\sec{s$^{-1}$}
\def\sm{$M_{\odot}$}
\def\x{$\times$}
\def\min{$'$\hspace*{-.1cm}}

\title{Stellar populations associated with the LMC Papillon Nebula\thanks
   {Based on observations obtained at the European Southern 
   Observatory, Paranal, Chile;  Program 66.C-0172(A). Table 1 is
   published only in electronic form.}
}

\offprints{Fr\'ed\'eric Meynadier, \hspace{1cm} \\Frederic.Meynadier@obspm.fr}

\date{Received 16 December 2003 / Accepted 22 March 2004}

\titlerunning{LMC N\,159}
\authorrunning{Meynadier et al.}

\author{F. Meynadier\inst{1} \and M. Heydari-Malayeri\inst{1}
 \and L. Deharveng\inst{2}  \and V. Charmandaris\inst{3,1} \and 
Th. Le Bertre\inst{1} \and \\M
.R. Rosa\inst{4,}\thanks
{Affiliated to the Space Telescope Division of the European Space Agency,
ESTEC, Noordwijk, Netherlands}
\and D. Schaerer\inst{5,6} 
\and H. Zinnecker\inst{7} 
}

\institute{{\sc lerma}, Observatoire de Paris, 61 Avenue de l'Observatoire, 
F-75014 Paris, France 
\and Observatoire
de Marseille, 2 Place Le Verrier, F-13248 Marseille Cedex 4, France
\and Cornell University, Astronomy Department,
106 Space Sciences Bldg., Ithaca, NY 14853, U.S.A. 
\and Space Telescope European Coordinating Facility, European Southern
Observatory, Karl-Schwarzschild-Strasse-2, D-85748 Garching bei
M\"unchen, Germany  
\and Observatoire de Gen\`eve, 51, Ch. des Maillettes, 
CH-1290 Sauverny,  Switzerland      
\and  Laboratoire d'Astrophysique, UMR 5572, Observatoire
 Midi-Pyr\'en\'ees, 14,  Avenue E. Belin, F-31400 Toulouse, France
\and Astrophysikalisches Institut
Potsdam, An der Sternwarte 16, D-14482 Potsdam, Germany 
}

\abstract{We study the Large Magellanic Cloud Papillon Nebula (N\,159-5), a 
conspicuous High Excitation Blob (HEB) lying in the star forming
complex N\,159. Using {\it JHK} near-infrared photometry obtained at
the ESO VLT with the ISAAC camera, we examine the stellar populations
associated with the Papillon, tracing their history using stellar
evolution models. Two populations are revealed: one composed of young,
massive stars with an age \ab\, 3 Myr, and a second consisting of
older lower mass stars of age spreading between 1 and 10 Gyr.  We
analyze the properties of those populations and discuss their
significance in the context of N\,159.  We also estimate that if the
star at the center of the Papillon is single its initial mass is
\ab\,50\,\sm\, and it is affected by an extinction $A_{V}$\,\ab\,7 mag. \\

 }

\maketitle


\begin{figure*}
\begin{center}
\centering\includegraphics[width = 19cm]{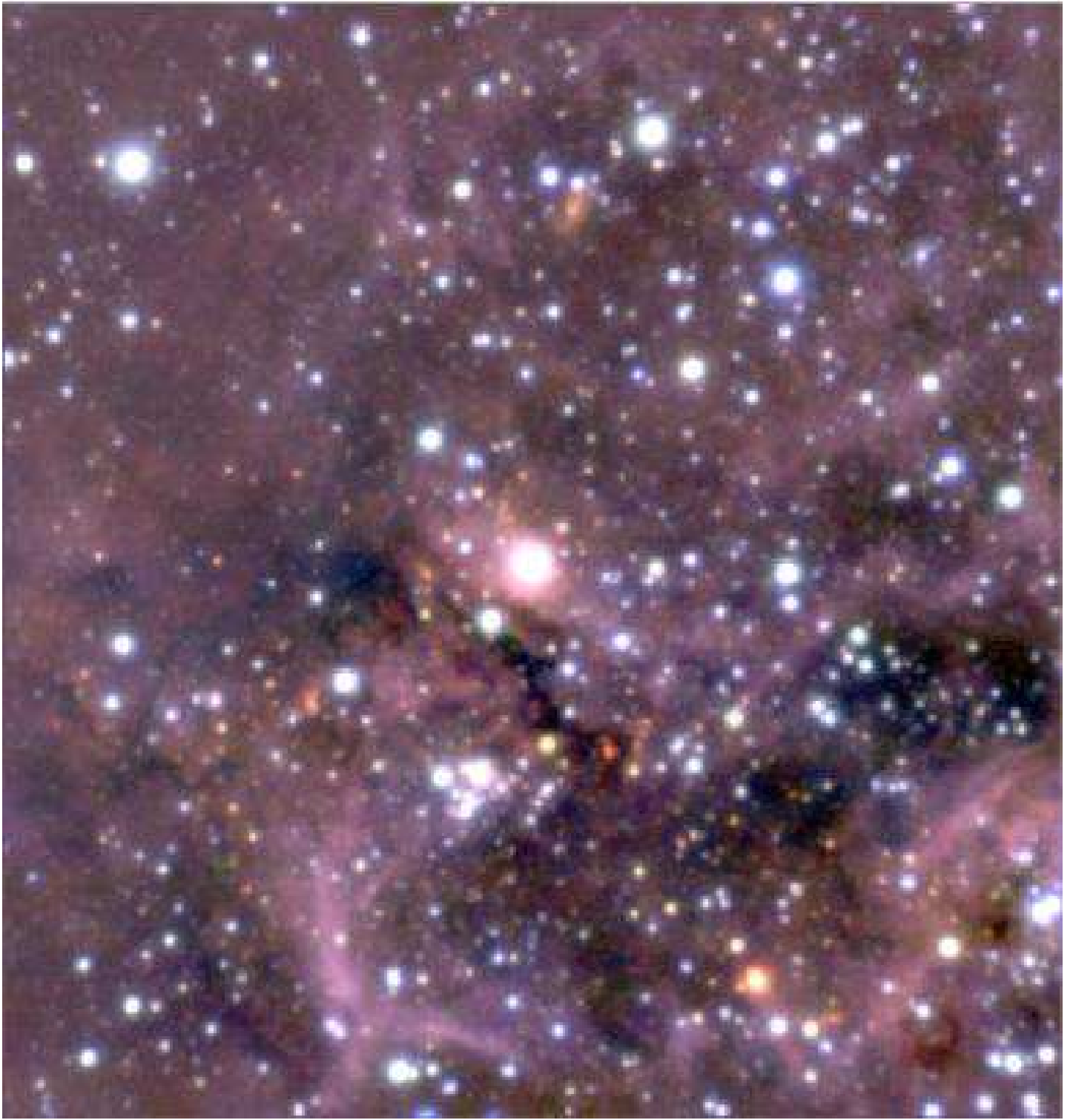}
\caption{{\it JHK} color composite image of LMC N\,159 
($Ks$ = red, $H$ = green, $J$ = blue) centered on N\,159-5,  the
Papillon nebula. North is up and East is left. See
Fig.\,\ref{chart_n159} for star identifications.  The field size is
2\min.1 $\times$ 2\min.2 (32\,pc $\times$ 33\,pc).
\label{jhk_n159} 
} 
\end{center}
\end{figure*}
\begin{figure*}[p]
\begin{center} 
\begin{minipage}[h]{\linewidth}
\centering\includegraphics[width = 19cm]{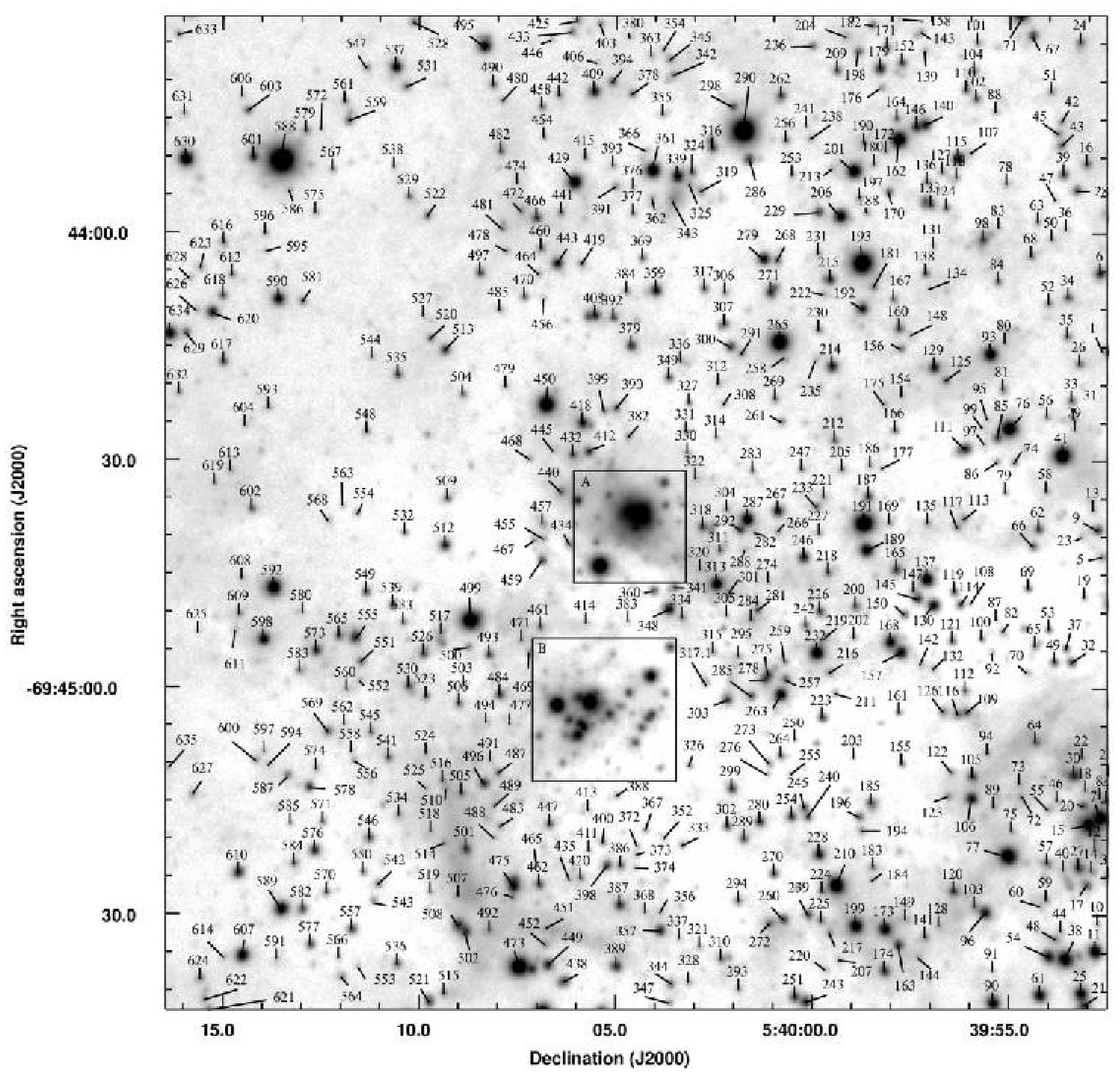}
\caption{Finding chart ($H$ filter) for the stars in Fig.\,\ref{jhk_n159}. 
The numbers refer to Table 1.  Regions A (N\,159-5) and B are detailed
below.
\label{chart_n159} 
} 
\end{minipage}
\end{center}

\begin{minipage}[h]{.60\linewidth}
\begin{minipage}[h]{.48\linewidth}
\centering\includegraphics[width = .9\linewidth]{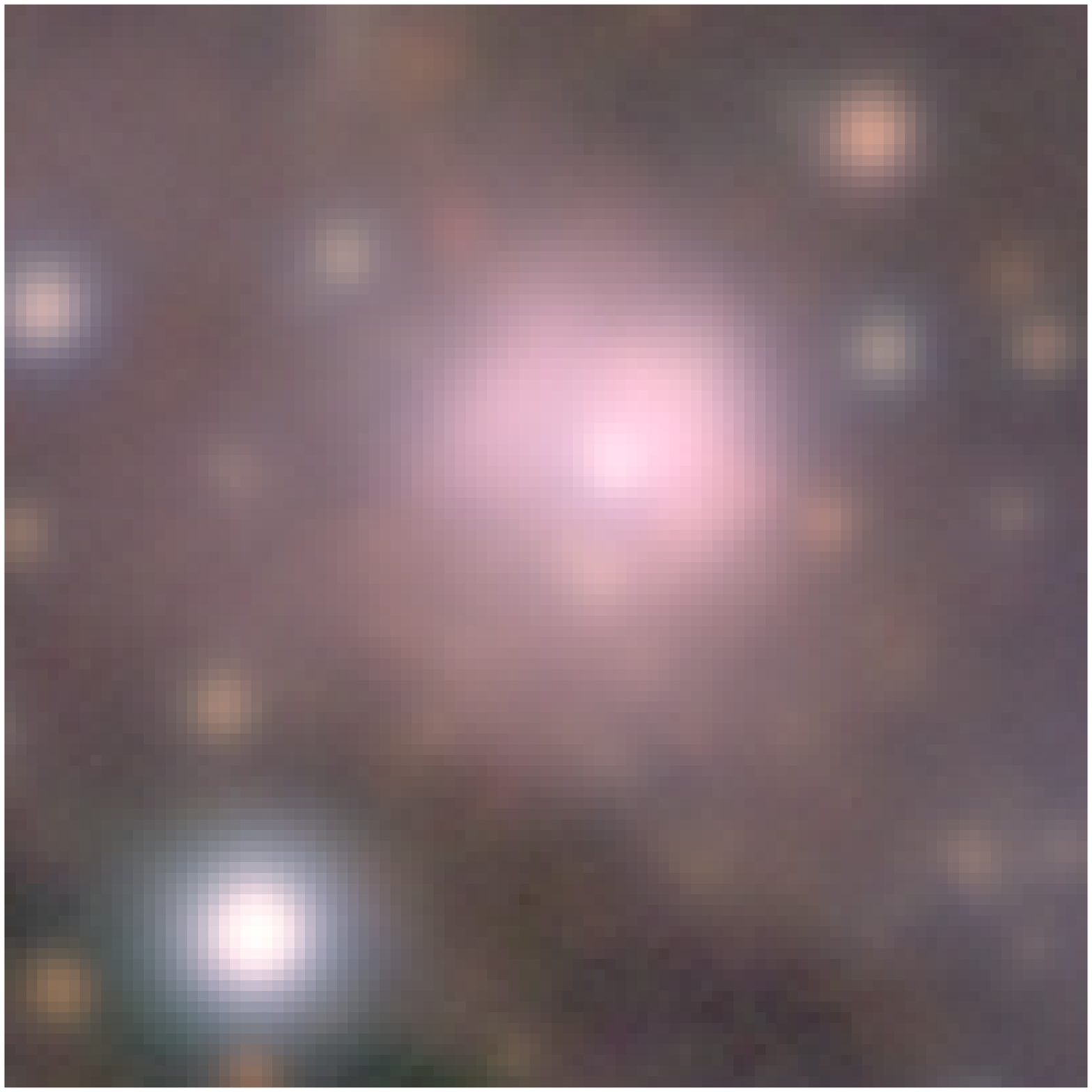}
\end{minipage}
\hfill
\begin{minipage}[h]{.48\linewidth}
\includegraphics[width = \linewidth]{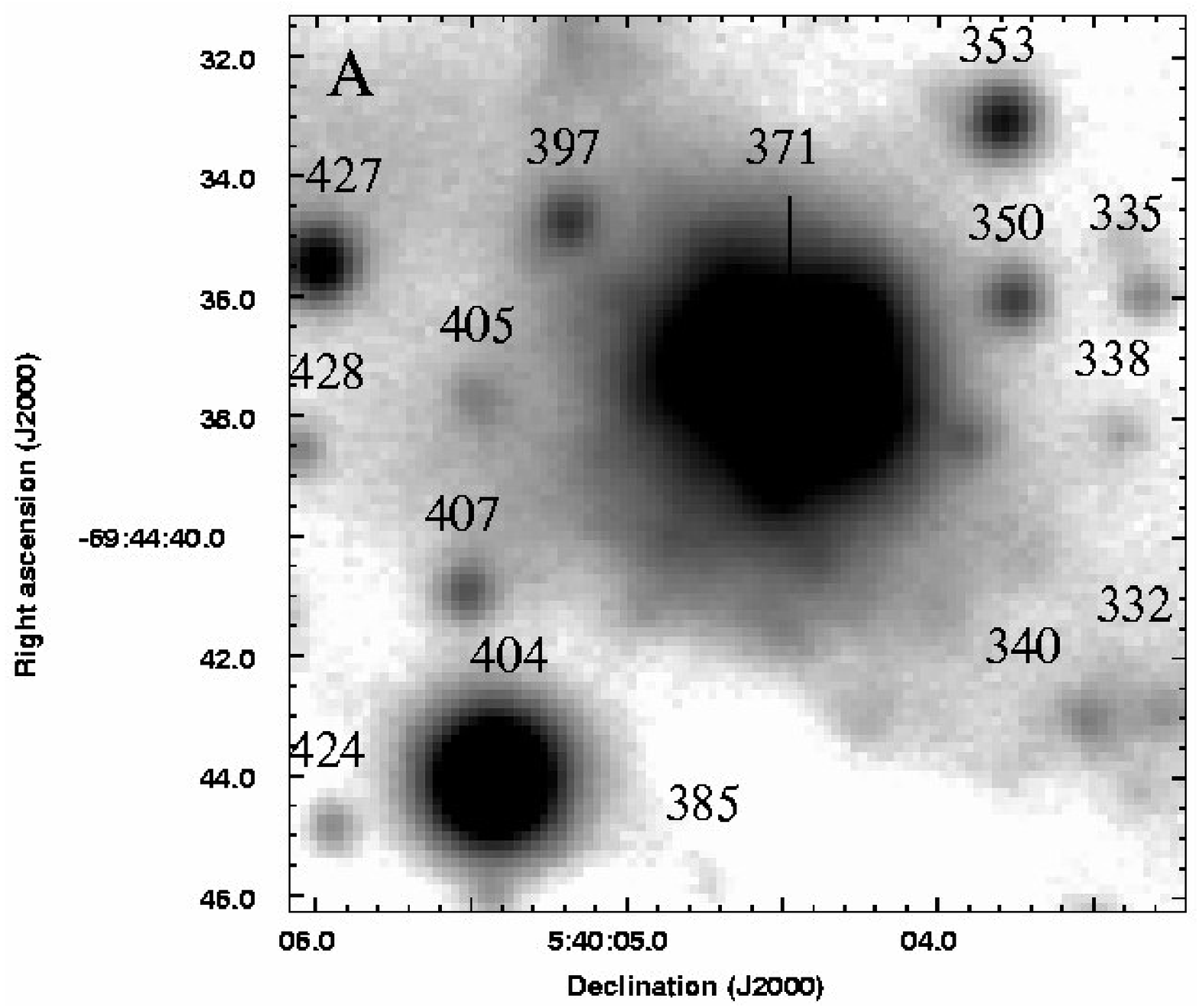}
\end{minipage}
\caption{
\label{chart_det}
Composite {\it JHKs} image ($Ks$ = red, $H$ = green, $J$ = blue) and
the corresponding finding chart ($H$ band) for region A, the central
Papillon (see Paper I).}
\end{minipage}
\hfill
\begin{minipage}[h]{.33\linewidth}
\includegraphics[width = \linewidth]{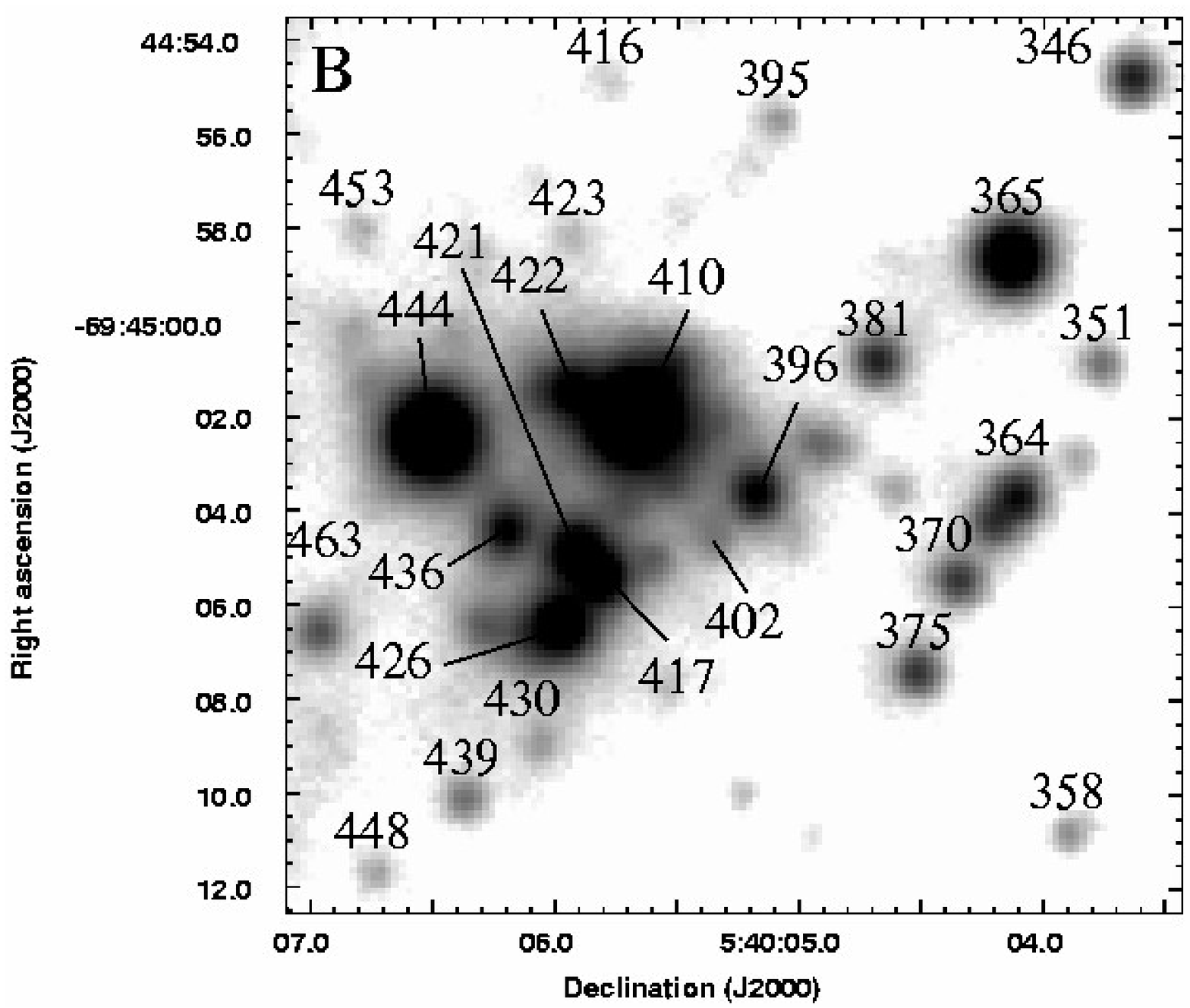}
\caption{Finding chart for region B ($H$ band).}
\end{minipage}
\end{figure*}

\section{Introduction}
The compact \h2 regions called High-Excitation Blobs (HEB) constitute
a rare class of ionized nebulae in the Magellanic Clouds.  They are
characterized by high excitation, small size, high density, and
large extinction compared to typical Magellanic Cloud \h2 regions.
These objects are tightly linked to the early stages of
massive star formation, when the stars begin to hatch from their
parental molecular clouds. For this reason their study
yields valuable information for a better understanding of 
the formation of massive stars.  \\

The \object{N\,159} complex \citep{henize} 
lies south of the famous starburst site 30 Dor and has attracted
special attention over the years. Its other designations are \object{MC\,77}
\citep{mcgee72}, \object{LH\,105} \citep{lh}, and \object{DEM\,271} 
\citep{dem}.  It
shows several signs of ongoing star formation activity, such as
infrared sources, cocoon stars, masers, and is also associated with
the most important concentration of molecular gas in the LMC
\citep{jones86,brooks97, johansson98}. The molecular emission is in
fact due to three distinct giant molecular clouds, known as N159-East,
N159-West, and N159-South.  Molecular lines tracing high density
regions are observed towards N\,159-W and N\,159-S in CS, CN, HCN, and
HCO$^{+}$ \citep{heikkila99}, while $^{13}$CO and upper-level
$^{12}$CO transitions, and the [\cii] emission line were mapped
towards the three giant molecular clouds, including N\,159-E
\citep{bolatto00}. The region we are interested in lies near 
N\,159-E. \\

The Papillon Nebula, also called \object{N\,159-5}, to which the present
study is devoted, is  the prototype of the HEB family    
\citep{mhm82}, which now possesses several members, such as N\,160A1,
N\,160A2, N\,83B, N\,11A in the LMC and N\,88A and N\,81 in the
SMC. Recent {\it HST} observations have resolved most of these objects
and revealed turbulent media typical of newborn massive star regions
marked by strong stellar winds interacting with the ambient ionized
gas \citep{mhm99a, mhm99b, mhm99c, mhm01a, mhm01b, mhm02a, mhm02b}.
These observations also showed a large extinction due to
local dust associated with ionized gas.  In a number of cases the
exciting sources were uncovered as a small cluster of massive stars.\\

In particular, the WFPC2 {\it HST} observations revealed that the
featureless blob N\,159-5 has in fact the morphology of a {\it
``papillon''}, i.e. it is a butterfly-shaped ionized nebula with the
``wings'' separated by \ab\,2\frac.3 (0.6 pc)
\citep[][ hereafter paper I]{mhm99c}.  Moreover, two subarcsecond features
resembling a ``smoke ring'' and a ``globule'' were detected in the
wings.  The images also showed a large number of subarcsecond
filaments, arcs, ridges, and fronts carved in the ionized gas by the
stellar winds from massive stars in the N\,159 complex. However, no
bright stars associated with the core of N\,159-5 could be identified
in the {\it HST} images.  Could this be due to the extinction by dust
large enough to hide the exciting stars? Since an $A_V \geq 6$\,mag was
needed to bring an O8 star below our {\it HST} sensitivity limit, we
decided to perform high resolution near-IR imaging of the region with
the ESO Very Large Telescope in order to address this issue and
explore the properties of the stellar population of N\,159.  \\


\section{Observations and data reduction}

\label{vlt}
The N\,159 region was observed in service mode with the ESO Very Large
Telescope (VLT). The infrared spectro-imager ISAAC was used at the
Nasmyth B focus of UT1 through filters {\it J} on 7 October and {\it H} and
{\it Ks} on 1st March 2001. The IR detector  (Hawaii Rockwell array) had
$1024\times1024$ pixels of 18.5\,$\mu$m each (0\frac.148 on the sky),
thus providing a field of 2\farcm5$^{2}$. \\

A set of individual, 10-second exposures was obtained in each filter
using the dithering method with a random offset of 15\frac\, at most.
The number of exposures were 10, 30, and 36 for the $J$, $H$, and {\it
Ks} bands respectively. The coadded frames have a spatial resolution
of 1\frac.08 for {\it J}, 0\frac.74 for {\it H}, and 0\frac.63 for
{\it Ks}.\\

PSF-fitting photometry was carried out in the $J$, $H$, and {\it Ks}
filters using the DAOPHOT II/ALLSTAR procedures \citep{stetson87}
under the ESO MIDAS reduction package. It should be noted that these
procedures are well adapted to the high-precision photometry of
globular clusters (i.e. tight groups of point sources with no
 background emission), but are not designed for handling regions with
very bright and variable background such as N\,159.  Some alternatives
to address those limitations have been investigated by
\citet{deharveng92}, and involve an iterative subtraction of the
background as derived from the approximative photometry obtained at
each step. In order to improve this method, we developed a software,
called DENEB for DE-NEBulized photometry. Our software enables an
interactive modification of the intermediate photometry files as well
as a real time check of the validity and convergence of those
modifications since it displays
the resulting residual background.\footnote{People
interested in the use or development of this tool are invited to
contact
\texttt{Frederic.Meynadier@obspm.fr}.} \\

Finally the frames were calibrated using the mean atmospheric
extinction coefficients and the color equations supplied by ESO, and
three standard stars for determining the zero points.   The
average photometric errors reported by DAOPHOT are 0.04, 0.04, and 0.05
mag for the faintest stars in $J$, $H$, and {\it Ks}  respectively.  The
relative accuracy is better than 0.01 mag for $J$, $H$, and {\it Ks}
brighter than 17 mag. \\

We compared the resulting magnitudes with those provided by the 2MASS
point source catalogue \citep{cutri03} using a selection of 36 stars
which appeared as single in our frames and were brighter than 15.0 mag
in {\it H}.   After correction for filter bandpasses
\citep{carpenter01, 2mass}
the mean differences are
m(2MASS)\,--\,m(ISAAC)\,=\,$-0.06$,  $-0.08$,  and
$-0.02$ mag in $J$, $H$, and {\it Ks} respectively. Taking into account 
the accuracy of the 2MASS photometry for stars of $H=
15$ mag  (\ab\,$\pm$ 0.1 mag r.m.s) and the uncertainties 
involved in  the
filter bandpass corrections, we considered these mean
differences not to be significant.\\

Our astrometry and image registration was
tied to the positions of the same 2MASS stars since it is known that
the r.m.s uncertainty in the positions of the 2MASS catalogue is $<$
0\frac.3. The astrometry and the photometry of the stars are given in
Table 1, which is available in electronic form as online material and
also at the Centre de Donn\'ees astronomiques de Strasbourg (via
anonymous ftp to cdsarc.u-strasbg.fr or via
http://cdsweb.u-strasbg.fr/Abstract.html).


\section{Results}

\subsection{Morphology}

A composite color {\it JHKs} image of the observed field is shown in
Fig.\,\ref{jhk_n159}, while the corresponding finding chart is
presented in Fig.\,\ref{chart_n159}. The stars are identified by a
number, according to Table 1.  Figs.\,\ref{chart_det} and 4 give
details on two densely populated regions, the central Papillon nebula
and a small southern cluster. \\

The field is fairly rich, with 896 stars detected at a $3\sigma$ level
in the $H$ band image (limiting magnitude 20), which has the best S/N
ratio. Among them 605 objects are detected at $3\sigma$ on all theses
frames.  Some particularly bright,
but highly reddened stars do not appear in all three filters and
consequently they were not included in the analysis. Exception was
made for a source labeled as \#317.1 which is not detected in the $J$
band while being relatively bright in {\it Ks}. \\

The image is marked by several dark regions and lanes indicating
strong absorption. The Papillon nebula is situated near the border of
a prominent central absorption lane.  The background is dominated by
ionized gas emission but is locally obscured by heavy extinction. The
southern edge of the field yields a fan-shaped filament already
visible on the {\it HST} frames (Paper I).  Our new {\it JHKs} imagery
provides a deeper overview of the stellar content of N\,159. We can
easily note that the small, bright cluster south to the Papillon (area
B of Fig.\,\ref{chart_n159}) is much more visible in the near IR 
than in the optical (Paper I).
The two reddest stars of the field are \#210 and \#317.1.  The first
one is located in the lower right quadrant of our images almost 1
arcmin from N\,159-5, and the second one near the edge of the
absorption lane. \\

We also note the presence of a number of ``peculiar'' objects having
an elongated form and a red color: \#343, \#517, and \#149.  The first
one seems also to have a tilted shape. The probability that they
result from a chance alignment of several faint red stars is very
low. Since they are physically too extended to be considered as
circumstellar disks of the LMC, it is more likely that they are simply
background galaxies.

\subsection{The infrared colors and stellar ages}

\begin{figure*}
\begin{center}
\includegraphics[width = \linewidth]{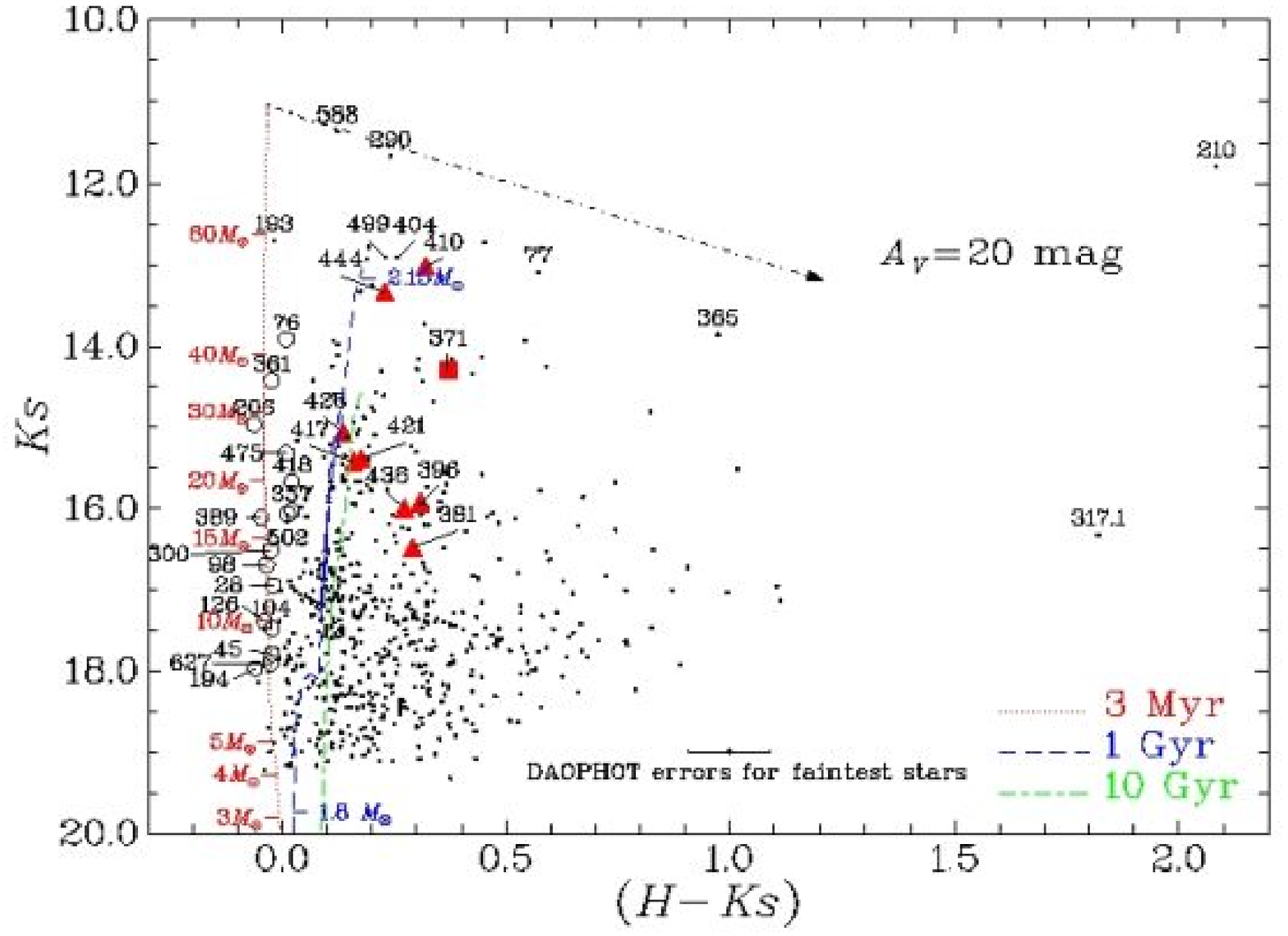}
\caption{
Color-magnitude, {\it Ks} versus {\it H -- Ks}, diagram for the
observed stars towards LMC N\,159 detected in all three
filters. Three isochrones are shown, 3 Myr (dotted red curve), 1 Gyr
(dashed blue), and 10 Gyr (dashed-dotted green), computed for a
metallicity of Z = 0.008 \citep{lejeune01} and a distance modulus of 
18.5 mag.  The upper and lower mass limits are indicated for the 1 Gyr
isochrone.  The reddening track, plotted as an arrow, extends to
$A_{V}$\,=\,20 mag.  The numbers refer to the stellar identifications
presented in Fig.\,\ref{chart_n159}.  Triangles represent a sample of
the stars belonging to region B, while the square (numbered \#371)
refers to the central point-like source of the Papillon.  Stars
brighter than {\it Ks} = 18 mag and situated near the 3 Myr isochrone
are labelled and shown as empty circles. 
\label{col-mag} 
} 
\end{center}
\end{figure*}

\begin{figure*}
\begin{center}
\includegraphics[width= \linewidth]{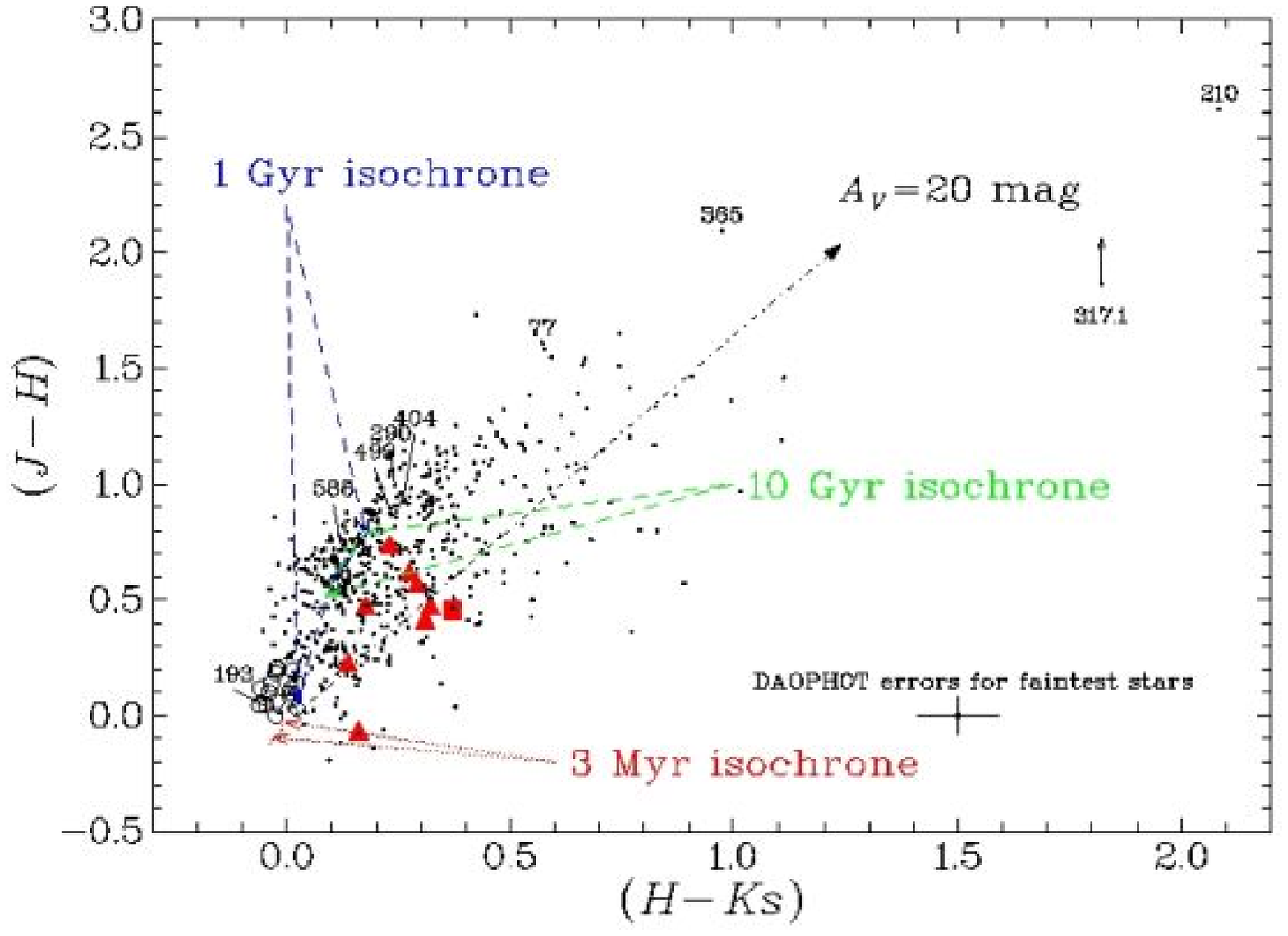}
\caption{
Color-color, {\it J -- H} versus {\it H -- Ks}, diagram for the
observed LMC N\,159 stars. The isochrone reference, 
symbols, and the reddening track are as in Fig.\,\ref{col-mag}. The
arrow attached to star  \#317.1 is due to the fact that an upper limit
of 20 in the $J$ band has been assumed for it.
\label{col-col} 
} 
\end{center}
\end{figure*}

Fig.\,\ref{col-mag} presents the {\it Ks} versus {\it H\,--\,Ks}
diagram for the measured stars towards N\,159, while
Fig.\,\ref{col-col} displays the corresponding color-color {\it J --
H} versus {\it H\,--\,Ks} diagram.  All sources brighter than {\it H}
= 20 mag and also detected in $J$ and {\it Ks} are taken into 
account.  Star \#317.1, which is not detected in $J$, is assigned an
upper limit of 20 in this band. We wish to point several stars in our
sample which display rather unique properties. The square in
Figs.\,\ref{col-mag} and \,\ref{col-col}, identifies star \#371, which
is the central point-like source of the Papillon. The eight triangles
correspond to the brightest components of the small southern cluster
marked as region B on Fig.\,\ref{chart_n159}.  \\


The color-magnitude and color-color 
 diagrams are interpreted by overplotting isochrones from
\citet{lejeune01} with  Z = 0.008 for a distance modulus of 
18.5 mag. As usual with near-IR observations, it is difficult to
discriminate low-mass old stars from young massive stars, because the
evolutionary tracks are nearly parallel to isochrones, resulting in a
very close location on the color-color diagram for those two
populations. This degeneracy is lifted if the mass is taken into
account: color-magnitude and color-color diagrams should be
``coherent'', in the sense that populations found to be fitted by a
given isochrone in one diagram should be fitted by the same isochrone,
within the same mass interval, in the second.  Uncertainties in the
photometry though, as well as lack of knowledge in the variation of
the extinction introduce limitations to the precise determination of
the corresponding isochrone. \\

The diagrams show the presence of two stellar populations. The first
one is a young population which appears to be fitted well with a 3 Myr
isochrone. Some of the members of this population are weakly affected
by extinction while other members have reddened colors.  The
extinction-free subset is vertically aligned around {\it H --
Ks}\,=\,0.00 in Fig.\,\ref{col-mag}, and the sample affected by
extinction has {\it H -- Ks} colors around 0.20 mag. This young
population is made up of massive O type stars, and may also contain a
component of reddened B type stars of \ab\,15\,\sm\, spread around
{\it H -- Ks}\,=\,0.2 mag.  Apart from this young population, there is
a second population with generally redder colors, which can be fitted
with much older isochrones of least 1 to 10 Gyr in age. 
The bulk of the stars in this population are fainter than {\it Ks} =
17 mag and have a mass of about  1 \sm,
although the brightest members have evolved into giants.
This population is also affected by a varying amount of extinction.
The points lying to the right of the 1-10 Gyr isochrones are much more
extincted, probably representing the stars situated deeper in the
molecular cloud. 
As was mentioned earlier,
it is not clear which precise isochrone should be used,
because we expect the extinction to be generally high and locally
variable. It is, however, 
evident that this second population is significantly older than the
first one, and we can notice the existence of a considerable
spread in age among this population.
It should also be underlined that for the metallicity of the LMC a
star of initial mass 2.15\,\sm\, evolves into a giant in less than
\ab\,1 Gyr. \\

The color-magnitude diagram can be used to estimate the extinction of
the stars. Assuming that the triangles represent young massive stars
of age \ab\,3 Myr, their shift to the right in Fig.\,\ref{col-mag} is
attributed to reddening by dust. An extinction of $A_{V}$\,\ab\,5 mag
is sufficient to displace the mean position of that stellar
population. The star \#371, detected towards the Papillon, does not
seem to have an IR excess, but is affected by an extinction of
$A_{V}$\,\ab\,7 mag.  This is consistent with paper I, which found
$A_V \geq 6$ for this central region on the assumption that extinction
would have to be large enough to hide an hypothetical O8 exciting
star.  See Sect. 3.3 for comparison with the CO map. \\

\subsubsection{Isochrone fitting}

\begin{figure}
\begin{center}
\includegraphics[width= \linewidth]{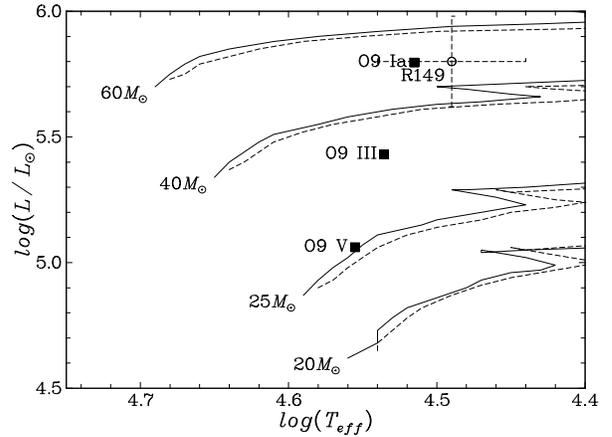}
\caption{
\label{L-Teff} 
Evolution tracks for several masses, plotted from the Geneva grid of
models \citep{lejeune01}. Solid lines: LMC metallicity, Z = 0.008;
dashed lines: Galactic metallicity, Z = 0.02.  Filled squares represent
the positions of O9 type stars of different luminosity class
\citep{vacca96}). Open circle shows the position of \#193 (R\,149),
assuming $M_{V}$ between $-6.8$ and $-6.6$, $T_{eff}$ between 27500
and 34300 K, and a bolometric correction between $-3.4$ and $-2.7$
\citep{vacca96}.  }
\end{center}
\end{figure}

\begin{figure}
\begin{center}
\includegraphics[width= \linewidth]{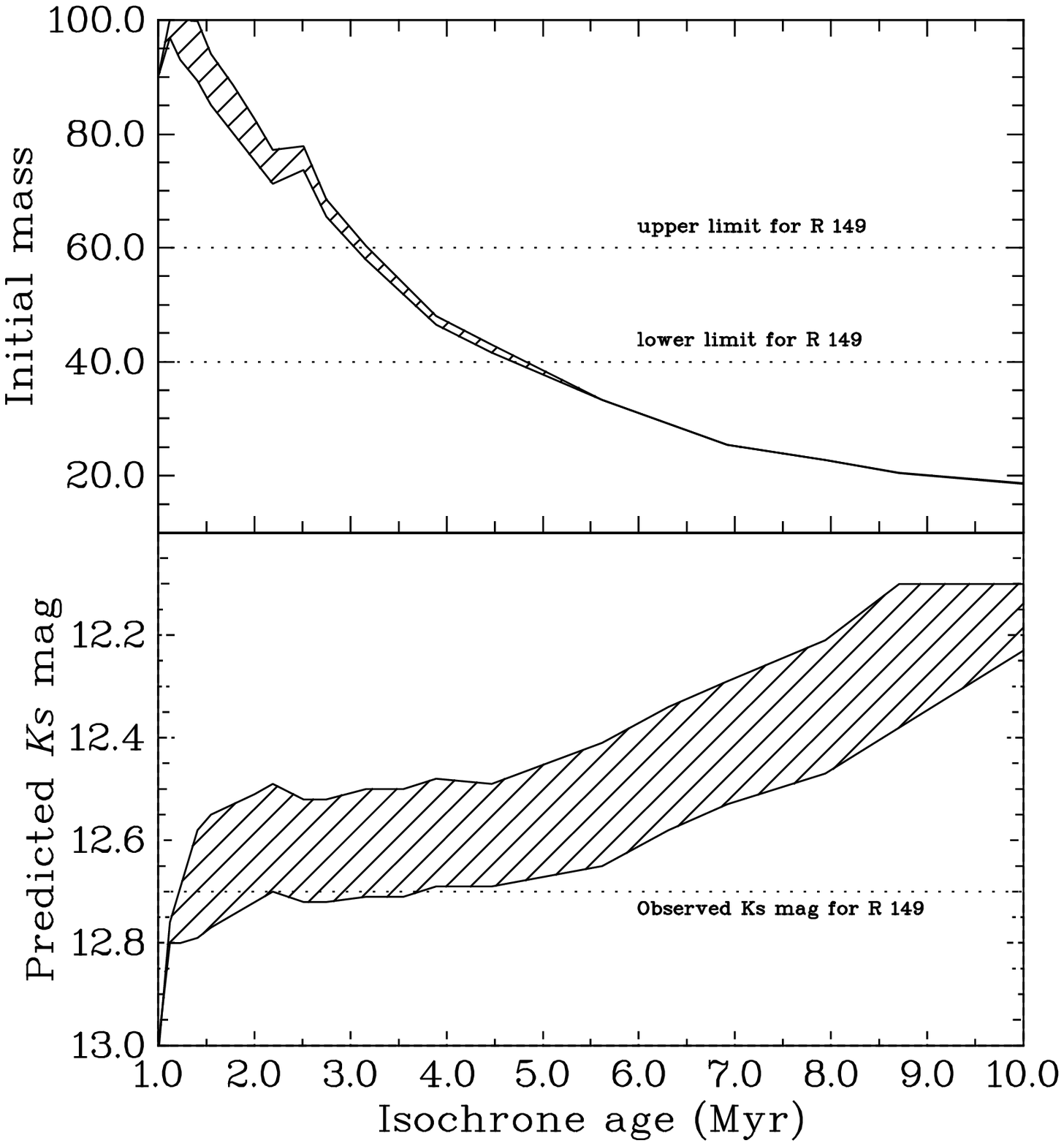}
\caption{
\label{fit} 
Expected initial mass (top) and {\it Ks} magnitude (bottom) for stars
with $M_V$ between $-6.8$ and $-6.6$ mag as a function of age. The
data is taken from the Geneva grid of models for Z = 0.008
\citep{lejeune01}.  Shaded areas correspond to points with $M_V$
between --6.8 and --6.6.  {\it Top}: The mass range deduced from
Fig.\,\ref{L-Teff} is represented by two horizontal dotted line at $M
= 60 M_\odot$ and $M = 40 M_\odot$.  {\it Bottom}: The observed {\it
Ks} magnitude is represented by a horizontal dotted line at {\it Ks} =
12.63 mag.}  
\end{center}
\end{figure}

The apparent magnitude of star \#193, better known as R\,149 or
Sk$-69^{\circ}$257, is $V=12.389$ with $B-V$\,=\,--0.081, 
$U-B$\,=\,--0.962 mag 
\citep{schmidt-kaler99}, in agreement with $V=12.49$ mag
\citep{dufour75}, and corresponds to an absolute magnitude of 
$M_V$\,\ab\,--6.8 to --6.6.   For an O9 spectral type \citep{walborn77, conti86}
this is marginally consistent with luminosity class III, but typical
of class I \citep{vacca96}. A dwarf classification, as suggested by
\citet{conti86}, seems therefore excluded.  The effective temperature
of O9\,III--I stars is $T_{\rm eff}$\,\ab\,34300--32700 K using
\citet{vacca96}'s calibration or typically between 31.6 and 27500 K
for the extreme Ia class based on recent model analysis taking into account
non-LTE line blanketing \citep{martins, crowther02, herrero02,
markova04}.  The bolometric correction, which is essentially
independent of line  blanketing, is then $BC$\,\ab\,$-3.4$ to $-2.7$
\citep{vacca96} translating to luminosities $\log L/L_\odot$\,\ab\,
5.6--6.0.\\

Fig.\,\ref{L-Teff} presents the Geneva evolutionary tracks calculated
for initial masses 20, 25, 40, and 60\,\sm\, and metallicities Z =
0.008 and 0.02 \citep{lejeune01}.  We reported the position of R\,149
on this diagram, using the values calculated in the previous
paragraph. It indicates an initial mass of the order of 40 to 60
M$_\odot$, which corresponds to an age of 3 to 5 Myr using the initial
masses predicted for stars of various ages and $M_V$ between --6.8 and
--6.6 mag (Fig.\,\ref{fit}, top). This is consistent with the age
range, 1 to 4 Myr, derived from the observed $Ks$ magnitude
(Fig.\,\ref{fit}, bottom).

\subsubsection{The brightest and reddest stars}

Based on the color-magnitude and color-color diagrams
(Figs.\,\ref{col-mag} \& \,\ref{col-col}),  stars \#588,
\#290, \#499, \#404, \#365, \#77, and \#210 may be high mass main
sequence members.  These stars can also be very tight multiple 
systems more or less affected by local dust. 
In particular, \#365 and \#210 are
very red, probably due to  their association with 
the prominent absorption  features 
in Fig.\,\ref{jhk_n159}. Furthermore, star
\#210 presents a near infrared excess typical of some Galactic 
OB exciting stars, for example star \#82 ionizing the \h2 region
Sh2-88B \citep{deharveng00}. It is not  easy to
estimate the extinction and the mass of  such a star,  
more especially since the possible 
presence of a circumstellar disk   alters  
the colors \citep{lada92}.
If single, this star would be one of the most 
massive stars of the region, having a mass of \ab\,100\,\sm\, while 
affected by an $A_{V}>$  20 mag.  
 \\

A number of reasons also suggest that 
some of these stars may be LMC supergiants. 
Using a bolometric correction of +2.7 mag in the 
$K$ band for supergiant stars \citep{vanloon99,lebertre01} and 
an extinction of $A_{K}$\,\ab\,0.5 or  1.0 mag, we find  
that the brightest stars of the sample have an
absolute   $M_{K}$ and 
bolometric $M_{b}$ magnitudes in the range 
--8 to --8.5 and  --5.7  to  --6.2  respectively.
These  magnitudes are consistent with  M type supergiants, carbon stars,   
or fainter AGB stars in the LMC \citep{vanloon99,vanloon00}. As 
for \#210, which has a redder color of {\it H -- Ks}\,=\,2.1 mag,  
it can qualify as an LMC AGB candidate. Future spectroscopic 
observations are needed in order to clarify the nature of these stars. \\

A third possibility is that at least some of these stars 
actually belong to our Galaxy and happen to be along the 
line of sight to the LMC.
We can make a rough estimate on their number by
establishing the H--R diagram of 2MASS sources found in a field
separated by a few degrees from the LMC. It appears that in our field
\ab\,14 sources brighter than {\it Ks} = 15 mag might be foreground
stars. Those stars cannot be compared to models computed for the LMC
distance modulus. Our observations also indicate that there are 45
stars brighter than {\it Ks = 15}, so approximately 30 of them should
be considered as belonging to the LMC.
The above-mentioned
bright stars have colors placing them  in the low-mass end of any isochrones 
between  1 to 10 Gyr (Fig.\,\ref{col-col}) adapted to the LMC, even though 
they are among the most luminous sources in Fig.\,\ref{col-mag} 
where they are located near the
high-mass end of the same isochrones. This apparent contradiction 
can be explained if they  are foreground Galactic stars: their
location should be compared to ``shifted'' isochrones in the
color-magnitude diagram in order to account for their different
distance moduli, while the color-color isochrones would remain
unchanged.   \\

\subsection{The molecular gas distribution}

\begin{figure*}
\begin{center}
\includegraphics[width = .7\linewidth]{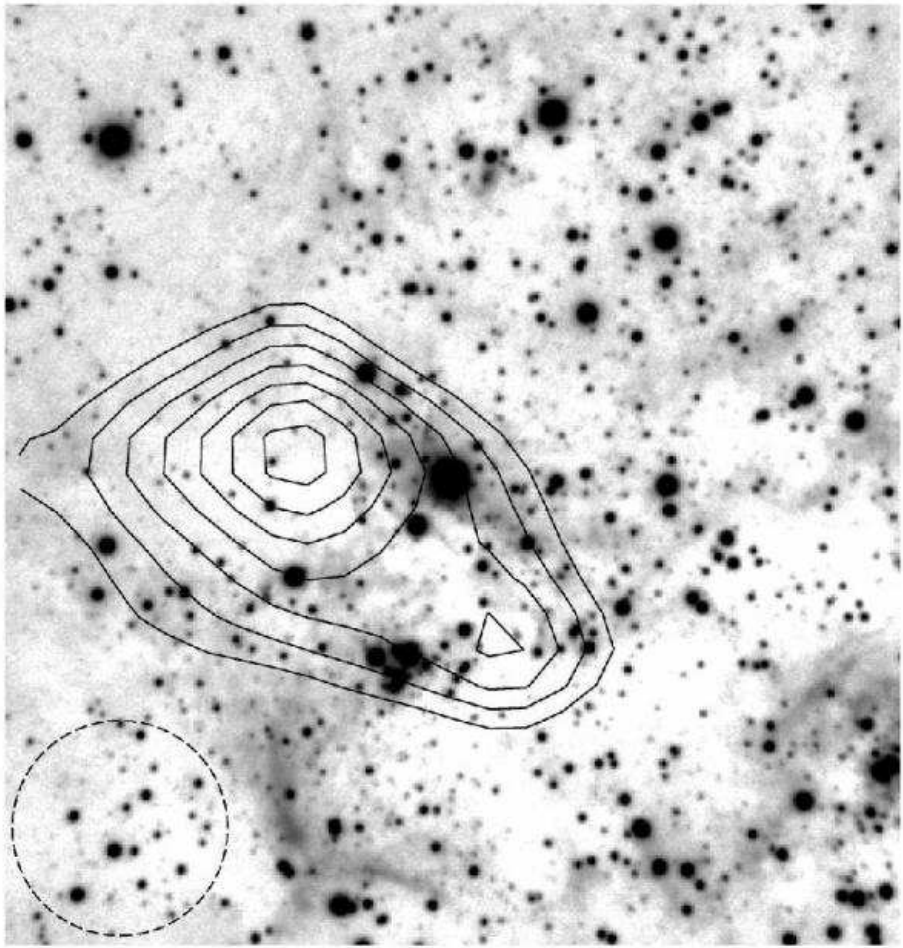}
\caption{
An $H$ band image of our field with a contour overlay of the 
\element[][12]{C}O(1--0) 
intensity of the molecular cloud N159-East from \citet{johansson98}. 
The field size and orientation are the same as in Fig. 1,
and the SEST \ab\,40\frac\, \element[][12]{C}O(1--0) beam is marked with a 
dotted circle.
\label{superpo} 
} 
\end{center}
\end{figure*}

\citet{johansson98} used the ESO SEST (Swedish European Submillimeter 
Telescope) to map the CO (1--0) emission towards N\,159 with a
resolution of 40\frac. We performed a bilinear interpolation between
each grid point in order to generate the contours corresponding to the
molecular gas associated with the Papillon region.  The result,
representing the CO emission peak called N\,159-East, is overlayed on
the $H$ image and presented in Fig.\,\ref{superpo}.\\

The constraints on $A_V$ established in paper I and in the present
work are in good agreement with Fig.\,\ref{superpo}. Since the mapping is
relatively scarce, the location of the two peaks cannot be precisely
determined but they coincide with the
absorption feature bordering the compact \h2 region. The present
picture is in perfect agreement with previous findings that the main
CO peak is shifted to the east of the bulk of the giant \h2 region N\,159
mapped in the radio continuum at 843 GHz
\citep{mills84,mhm85,israel96}.  Regions A and B are both adjacent to
the molecular peaks, region B being less affected by extinction.  It
is conceivable that more massive stars be in the process of birth
towards the CO emission maxima. \\

In order to estimate the extinction corresponding to the CO peak the
optically thin $^{13}$CO transition is needed. \citet{bolatto00}
observed the N\,159-W component in $^{13}$CO and derived a column
density of 1.1\,\x\,10$^{22}$ cm$^{-2}$ for the molecular hydrogen
H$_{2}$, corresponding to a column density of atomic hydrogen of
2.2\,\x\,10$^{22}$ cm$^{-2}$.  It is known that the gas-to-dust ratio
$N_{H}/E(B-V)$ in the LMC is several times larger than the Galactic
value \citep{nandy81,clayton85,lequeux89}. Using the conversion
relation $N_{H}/E(B-V)$\,=\,2\,\x\,10$^{22}$ atoms cm$^{-2}$
mag$^{-1}$ given by \citet{lequeux89} and $R=A_{V}/E(B-V) = 3.1$, we
find a visual extinction of $A_{V}$\,\ab\,3.5 mag for N\,159-W.  The
extinction for N\,159-E should be smaller since we know that N\,159-E
is less dense than N\,159-W (see below).  On the other hand,
\citet{dickey94} carried out 21-cm \hi\, absorption line observations
against background continuum sources towards N\,159 using the
Australia Telescope Compact Array (ATCA) interferometer. Their \hi\,
cloud 0539--697 can be identified with the molecular cloud N\,159-E,
based on velocity similarity \citep{johansson98}. The CO cloud has
the following characteristics: $V$ = 238 km s$^{-1}$, $\Delta V$ = 6.0
km s$^{-1}$, log{\it (L$_{CO}$)} = 4.28 K km s$^{-1}$ pc$^{2}$,
whereas those for the \hi\, cloud are: $V$ = 244 km
s$^{-1}$, $\Delta V$ = 1.6 km s$^{-1}$, N(\hi) = 4.46\,\x\,10$^{22}$
cm$^{2}$. This column density corresponds to a visual extinction of
$A_{V}$\,\ab\,7 mag while that for N\,159-W, i.e. N(\hi) =
9.62\,\x\,10$^{22}$ cm$^{-2}$, indicates a larger extinction of
$A_{V}$\,\ab\,15 mag. A reason why
\hi\, observations yield stronger extinctions is  that the
higher spatial resolution of the interferometer picks up denser
clumps, in contrast to CO observations which are affected by beam
dilution. Moreover, it is quite possible that both techniques do not
exactly sample the same zones. Anyhow, the higher values are supported
by our {\it HST} (Paper I) and present ISAAC observations. \\

\section{Discussion}

The population of young massive stars, which was discussed earlier
(Sect. 3.2), is spatially distributed over the whole field, while a
sample of it,  represented by triangles in Figs.\,\ref{col-mag} and
\ref{col-col}, is  grouped in a cluster, marked as region B. 
This grouping is expected given the young age of these stars.
How can though one explain the separation between this cluster and the
other massive stars which even if they have similar ages are at a
considerable distances from the cluster, for example
\ab\,70\frac\, (\ab\,18\,pc) for 
star \#193, one of the most distant?  One
explanation could be that massive star formation may have taken place
simultaneously at different parts of the molecular cloud.  At these
locations the molecular material has been fully dissociated and
ionized, and we do not observe it now.  Alternatively, massive star
formation   has  occurred in cluster B, and subsequently a number of the
members have been ejected due to the dynamics of the cluster.\\

It has been shown that once an embedded cluster forms, three mass
evacuation mechanisms work over different timescales to
disrupt it
\citep{kroupa01}: $a)$ expulsion of embryonic gas (approximately during
the first 0--5 Myr), $b)$ mass loss from evolving stars (significant after
\ab\,3 Myr), and $c)$ stellar dynamical evaporation and ejection of
stars (all times).  Binary-binary collisions are required to produce
high velocity escapees to occur in low density clusters
\citep{leonard88,leonard90}, although simple calculations 
suggest that such interactions are rather unlikely.  Recently
\citet{vine03} have studied the dynamics of massive stars in young
clusters containing gas and stars. They have shown that the location
of massive stars outside the core of the cluster does not exclude
their formation in the dense cluster core. The massive stars could
have originated in the core, but escaped from that region during the
gas expulsion phase.  Furthermore, the ejection of the OB stars must
have happened during an earlier evolutionary stage when the cluster
was most probably more compact than today
\citep{portegies99}. Assuming that star \#193 has been kicked out of
cluster B, an escape velocity of \ab\,5.5\,km\,s$^{-1}$ has been
necessary for it to reach its observed position after a travel time of
3 Myr. This estimate is a lower limit due to projection 
on the sky of a three-dimensional configuration in space.  Higher
velocities are quite possible since escapees can leave their
birthplace with velocities up to 100 km\,s$^{-1}$ or even larger
\citep{leonard90,kroupa95}.  \\

We note also that all the candidate massive stars are devoid of proper
nebulosity, in contrast to the Papillon. This fact suggests that the
Papillon is probably the youngest visible massive star formation event
in the whole field. The strength of the molecular hydrogen emission
detected towards the Papillon confirms its nature as a very young
star formation region \citep{israel88,kawara88}. In fact the observed
luminosity of the H$_{2}$ line {\it v = 1 -- 0 S(1)} towards the Papillon is
two times larger than that observed at the Orion source
\citep{kawara88}. The massive star(s) powering the Papillon have not had
enough time to disrupt the \h2 region.  Moreover, the presence of
nebulosity excludes the possibility for the Papillon of  ejection from
cluster B. It is therefore conceivable that the Papillon lies somewhat
above or below the mean plane of N\,159.
We believe that the
Papillon is situated at the side nearer to us since it is visible in
the optical.
We estimate that star \#371, which lies towards the center of the
Papillon, has a mass of \ab\,50\,M$_\odot$, even though based on our
current resolution we cannot exclude the possibility that the star is
multiple.  Should other massive stars be embedded inside that nebula,
much better spatial resolution and deeper exposures are required in
order to uncover them. From our previous {\it HST} observations we
estimated an exciting star of type at least O8\,V, \ab\,30 \sm\,
\citep{vacca96}, for the Papillon using the \hb\, flux measurement  
and assuming that the \h2 region is ionization bounded (Paper I). The
difference between the two mass estimates is due to the fact the \h2
region is density-bounded at least towards us and that the flux
correction for extinction is not straightforward. The latter point is
probably the reason why the radio continuum observations, which are
less affected by extinction, yield a higher Lyman continuum flux.  In
an earlier work \citep{mhm85}, we used the radio continuum
observations at 843 MHz, obtained with a beam of 43\frac\,\x\,46\frac,
to derive a flux density of 55 mJy for N\,159-5, after correcting for
contamination by the surrounding field.  A resulting ultraviolet flux
of \ab\,1.2\,\x\,10$^{49}$ photons indicates an O7\,V type star of
\ab\,38\,\sm\, \citep{vacca96}.  Given the uncertainties involved,
the stellar mass estimates based on the \hi\, emission from the nebula
agree well with the \ab\,50 \sm\, derived from photometry using
evolutionary models.  \\

An age of \ab\,3 Myr was derived for the massive star population using
the evolutionary models and supplementary data on one of the members.
We wish to note though that this age estimate may not be very accurate
due to the degeneracy of the near IR colors of massive stars. In fact
any isochrone between 1 and 10 Myrs would be consistent with our data.
We favored the 3 Myr isochrone in order to meet the requirements of
star Sk$-69^{\circ}257$.
\\

One can also estimate the number of stars which power the \h2 region
N\,159 on the basis of radio continuum observations. \citet{clarke76}
measured a radio continuum flux density of 6.5 Jy at 408 MHz using the
Molonglo telescope whose beam had a width of 2\min.6\,\x\,2\min.9. The
beamwidth is comparable with the size of our ISAAC field, and the
target coordinates match  the position of the Papillon, while
the reported pointing accuracy is 18\frac\,\x\,5\frac. The derived
Lyman continuum flux of 1.36\,\x\,10$^{51}$ photons s$^{-1}$
corresponds to some 40 massive stars of type O5\,V with an initial
mass of \ab\,50\,\sm\,\citep{vacca96}. Taking a Salpeter-like initial
mass function with slope $ x = - 1.5$, we can predict the presence of
some  360 stars of mass about 10\,\sm\, and  3240 stars 
of \ab\,2\,\sm. Where are these 40 O5\,V stars? 
It is quite possible that they are among the stars we imaged but due
to the degeneracies in the colors mentioned earlier they can only be
clearly identified if spectroscopic observations were available. 
Moreover, some of them may be embedded in the molecular cloud and 
some situated outside our ISAAC field.\\

The color-mag diagram (Fig.\,\ref{col-mag}) also shows the presence of
intermediate mass stars of \ab\,4--10\,\sm\, on the main sequence
formed together with high mass stars \ab\,3 Myr ago. This is in
agreement with more detailed results on the Orion Nebular Cluster 
(ONC) based on a large body of data  (\ab\,3500 stars
identified within \ab\,2.5 pc of the Trapezium, among which at least
\ab\,1600 with photometric and spectroscopic data in the visible) 
\citep{hillenbrand97}.  According to these studies,
low-, intermediate-, and high-mass stars have formed together in the
ONC a few Myr ago \citep{palla99}.
However, this may not be a universal trend since
\citet{herbst82}'s study of NGC\,3293 led them to the
conclusion that in a cluster low- and intermediate-mass stars form
first, with the process continuing gradually until the high-mass stars
appear. This result is in agreement with more recent findings on star
formation in LMC clusters and associations. For
instance, in the case of the R\,136 cluster, situated in the LMC 30
Dor, \citet{massey98} arrived to  the conclusion that intermediate-mass
stars began forming some 6 Myr ago and continued up to the time when
the high-mass stars formed, 1--2 Myr ago.  \\

An interesting question is  whether the young (\ab\,3 Myr) and
old (\ab\,1--10 Gy) stellar populations have formed in the same region
of space.  Although presently we do not have the necessary data to 
address this issue and cannot reach a firm conclusion, 
it is quite possible that both populations be spatially
unrelated. The LMC is known to have a considerable depth, the old
population can have formed in a different depth 
during much earlier star formation activities.  In order to get some
insight about this question, we used the 2MASS data to probe a bare
stellar field devoid of any particular nebular emission lying near the
N\,159 complex (radius 1\min.22, centered on  $\alpha$ = 05h\,39m\,00s,
$\delta = -69^{\circ}$\,47\min\,30\frac\,).  The corresponding HR
diagram shows the absence of a young, unreddened population, but the
presence of an old population resembling the one found towards
N\,159. Although this population is relatively smaller in number with
respect to that of N\,159, since 2MASS is not as deep as  our 
photometry, the presence of the old population is certain. The old
population seems therefore to be a common background stellar component
towards this part of the LMC.\\

The presence of low-mass pre-main sequence LMC stars in the above
diagrams seems unlikely, even if those objects are characterized by
large near IR colors, {\it H -- K}\,\ab\,1.5 mag
\citep{lada92,chabrier00}. A pre-main sequence star of \ab\,1 \sm\, 
has a luminosity of  log {\it (L/$L_{\odot})$}\,\ab\,1 on its
birthline, corresponding to an observed visual magnitude of \ab\,21,
which is below our detection limit. An intermediate mass
pre-main sequence star of 5\,\sm\, has an effective temperature of
\ab\,11,000 K and log {\it (L/$L_{\odot})$}\,\ab\,3, corresponding to
a reddened 16 magnitude star, 
 occupying loci around {\it J -- H}\,\ab\,0.5 and 
{\it H -- K}\,=\,0.5 mag \citep{lada92}.
There is  a few  number of sources with such colors in 
Fig.\,\ref{col-col}, given the color uncertainties at those magnitudes.
Therefore, we cannot exclude the possibility that some of those points
represent intermediate mass PMS stars.
As for more massive objects, the concept of pre-main sequence is 
not applicable to stars above \ab\,6\,\sm\, since the birthline and 
the ZAMS unify at those mass levels \citep{palla93}. \\

Comparison between LMC N\,159 and SMC N\,81 points out dramatic
differences between the environments of these two HEBs. The present
work shows the Papillon as part of a rich complex containing a large
molecular cloud and a cluster of young, massive stars, whereas our
previous study of SMC N\,81, based on ISAAC near IR observations
\citep{mhm03}, revealed a solitary star formation event.  Moreover, 
since the two compact \h2 regions  
have several comparable 
characteristics,  if we assume  that they have gone through a similar
formation process, then the HEB formation 
can take place in both very 
dense as well as  rather sparse environments. \\


\acknowledgements{We are grateful to Dr. L.E.B. Johansson for
  providing us with the CO map of the N\,159 molecular cloud.  VC
  would like to acknowledge the support of JPL contract
  960803. FM wishes to thank Dr. Eric Mandel for his valuable
  help concerning the DS9 astronomical data visualization application
\citep{ds9}. We would like also to thank the
  referee, Dr. Joao Alves, for useful advices.  Finally, this
  publication makes use of data products from the Two Micron All Sky
  Survey, which is a joint project of the University of Massachusetts
  and the Infrared Processing and Analysis Center/California Institute
  of Technology, funded by the National Aeronautics and Space
  Administration and the National Science Foundation.  }

\bibliographystyle{aa}
\bibliography{meynadier}

\Online
\onecolumn

\tablecaption{Astrometry and photometry of sources, with DAOPHOT errors}

\tablefirsthead{%
\hline
\hline
ID &  RA (J\,2000.0)& DEC (J\,2000.0)& $J$ (mag) & $H$ (mag) & $Ks$
(mag) 
& $\sigma J$ 
& $\sigma H$ 
& $\sigma Ks$ 
& notes\\
\hline
}

\tablehead{ \multicolumn{10}{l}{\textbf{Table 1.} continued...}\\
\hline
\hline
ID &  RA (J\,2000.0)& DEC (J\,2000.0)& $J$ (mag) & $H$ (mag) & $Ks$
(mag) 
& $\sigma J$ 
& $\sigma H$ 
& $\sigma Ks$ 
& notes\\
\hline
}

\tabletail{
\hline
\multicolumn{10}{r}{continued...}\\
}

\tablelasttail{
\hline\\
}

\begin{supertabular}{lllllllllp{2cm}} 
\label{table1}
  3     & 05:39:52.52   & -69:45:24.9   & 18.32         & 18.31         & 18.18         & 0.019         & 0.031         & 0.031 &\\ 
  4     & 05:39:52.55   & -69:45:14.9   & 17.54         & 17.26         & 17.17         & 0.009         & 0.018         & 0.012 &\\ 
  5     & 05:39:52.66   & -69:44:42.7   & 18.50         & 17.83         & 17.56         & 0.014         & 0.011         & 0.011 &\\ 
  6     & 05:39:52.71   & -69:44:05.3   & 16.04         & 16.03         & 15.99         & 0.005         & 0.003         & 0.004 &\\ 
  7     & 05:39:52.71   & -69:45:17.2   & 14.22         & 14.04         & 13.93         & 0.004         & 0.005         & 0.004 &\\ 
  8     & 05:39:52.71   & -69:45:14.6   & 19.03         & 18.65         & 18.38         & 0.038         & 0.040         & 0.041 &\\ 
  9     & 05:39:52.77   & -69:44:39.1   & 17.45         & 16.53         & 16.21         & 0.007         & 0.006         & 0.006 &\\ 
 10     & 05:39:52.77   & -69:45:31.2   & 18.90         & 18.31         & 18.14         & 0.017         & 0.020         & 0.021 &\\
\end{supertabular}
\end{document}